# HyLine: a Simple and Practical Flow Scheduling for Commodity Datacenters

Soheil Abbasloo, Yang Xu, H. Jonathan Chao ({ab.soheil, yang, chao}@nyu.edu)
New York University

*Abstract*— Today's datacenter networks (DCNs) have been built upon multipath topologies where each path contains multiple links. However, flow scheduling schemes proposed to minimize flow completion times (FCT) in DCNs are based on algorithms which are optimum or close-to-optimum only over single link. Moreover, most of these scheduling schemes seek either fully centralized approaches having overhead of communicating to a central entity or fully distributed approaches requiring changes in the fabric.

Motivated by these shortcomings, we present HyLine a simple scheduling design for commodity DCNs which is equipped with a joint load-balancing and flow scheduling (path-aware) design exploiting the multipath nature of DCNs. HyLine takes a hybrid approach and uses the global-awareness of centralized and agility of distributed techniques without requiring any changes in the fabric. To that end, it determines a threshold margin identifying flows for which using centralized approach is beneficial.

We have shown through extensive ns2 simulations that despite HyLine's simplicity, it significantly outperforms existing schemes and achieves lower average and 99th percentile FCTs. For instance, compared to Qjump–state-of-the-art practical scheme– and pFabric–one of the best performing flow scheduling schemes– HyLine reduces average FCT up to 68% and 31%, respectively, under a production datacenter workload.

## I. INTRODUCTION

User satisfaction (and total revenue) of today's popular datacenter applications such as search, social networks, and recommendation systems is closely related to the response times of these interactive applications. This motivates recent research to propose new datacenter (DC) transport designs for minimizing average flow completion times (AFCT) as the primary objective that is mainly determined by the end-to-end latency of datacenter networks (DCNs).

Prioritization is one of the main techniques used by different approaches to achieve lower AFCTs [1-5]. Wide range of these proposals use shortest remaining processing time (SRPT) (or its simplified versions), the optimum scheduling algorithm when used over a single link [1], to minimize AFCT in DCNs. However, as we show in section III, these algorithms are suboptimal for minimizing AFCT when each path in the network has multiple links. This issue will be escalated when multipath nature of today's DCNs is considered.

Agility of fully in-network schemes motivates some proposals to keep all changes in the network to achieve lower response times [1, 6, 2]. However, this usually requires changes in the fabric which brings extra costs for the datacenter owners [18, 7]. On the other hand, using centralized schemes such as [8], in which fabric will not be modified, comes at cost of performance degradation due to the delay introduced by the controller. This will be escalated when it is considered that most of the DC flows are very small and can be finished in just a few round trip times (RTTs) [9, 7]. Moreover, using explicit rate control mechanisms to precisely adjust flows' rates in the network leads to high complexity in the centralized approach (e.g., [3]) or the need to modify switches to coordinate with each other for finding and maintaining the best rates in the distributed approach (e.g., [6, 2]).

To overcome these shortcomings, in this paper, we present HyLine, a simple and practical flow scheduling design which:

1. Takes a hybrid approach requiring no changes in the fabric, and uses both global-awareness of centralized and agility of distributed techniques such as priority flow control (PFC) in layer 2,

2. Uses a joint load-balancing and flow scheduling (path-aware scheduling) policy to exploit the multipath nature of DCNs, and

3. Does not use any complicated per flow rate adjustment mechanism.

To that end, HyLine determines a threshold identifying 2 categories of flows: flows that should be scheduled in a centralized manner (2nd class flows having sizes larger than the threshold) and flows that should not be (1st class flows having size smaller than the threshold). Having that threshold, end-hosts simply assign $1^{st}$ class flows to the higher priority queue in commodity switches and send them to the network at line rate (TCP handles any further required rate adjustment). $2^{nd}$ class flows that are assigned to the lower priority queue will be scheduled before coming to the network. Each of the 2nd class flows should first send a request including flows' information to HyLine's central MANager (MAN) seeking its permission. MAN is responsible to control 2nd class flows in a very simple stop-and-go fashion. To do that, it uses simple path-aware scheduling policy to find the best path for the requested flow based on flow's information (priority). If a path is found for the new flow, MAN sends back a Go signal carrying the path that should be used by the corresponding flow. All permitted 2nd class flows enter the network at their end-host's line rate using the assigned paths (each edge-link will be used by at most one 2nd class flow at a time). MAN also sends a Stop signal to the preempted flows or the ones that cannot be served yet.

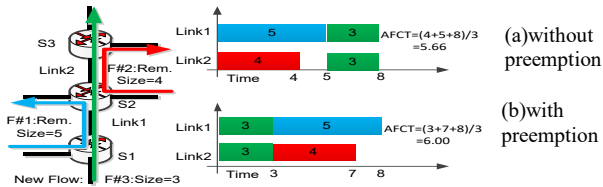

Fig. 1. AFCT with and without preemption.

We evaluate HyLine's performance through extensive packet-level simulations in ns2 [10]. The results show that despite simple nature of HyLine's design, it significantly outperforms recent schemes including pFabric [1], one of the best performing flow scheduling schemes, Qjump [5], the state-of-the-art practical scheme, and DCTCP [9]. In particular, compared to pFabric, Qjump, and DCTCP, HyLine reduces AFCT up to 31%, 68%, and 88% respectively, under a realistic DCN workload [9].

## II. RELATED WORKS

**Transport Designs:** There are vast number of TCP designs targeting a specific environment (e.g., [28] in cellular context and [3] in DCN context). Most of recent TCP proposals in DCN context use various prioritization mechanisms to minimize FCT [1, 4, 2, 3]. For instance, they assign different rates to flows based on their criticality [2], tag each packet with its corresponding priority and serve it regarding that priority in the network [1], use strict priority scheduling among queues in switches and assign flows dynamically to different levels of priority [4], or use a combination of these strategies [3]. Although designs that use the prioritization idea achieve good performance, they all are based on single-path scheduling algorithms such as SRPT. Therefore, some of these schemes (e.g., [3, 24]) only test their designs in single-path scenarios. Most of the other ones including [1, 4, 11, 5] use packet spraying [12] as load balancing mechanism to run their schemes on a multipath DCN. However, packet spraying is not an available feature in most of the commodity switches and is not used in commodity DCNs [18, 7, 13]. Therefore, we avoid using such load balancing mechanisms in this paper, though they might lead to good performance.

**Joint Transport-Load Balancing Designs:** Almost none of the load-balancing designs in the network layer are priority-aware. To the best of our knowledge, there is only one scheme called DeTail [14] in which a cross-layer approach is used to reduce the long tail of FCTs in DCNs. Although DeTail achieves good performance, a lack of backward compatibility and the need for changing both switches and end-hosts make it very hard if not impractical for commodity DCNs. Fastpass [8] uses a centralized entity to handle not only scheduling block but also load balancing block. However, it also follows the traditional approach of designing scheduling block (timeslot allocation block in [8]) and load balancing block (path selection block in [8]) separately. Moreover, Fastpass could not minimize FCTs, because at least for the very small flows that could be finished in a few RTTs, it adds (at least) one RTT delay caused by communication with Fastpass's central controller.

**Load Balancing Designs:** Nearly all load balancing schemes in DCNs are designed based on the fairness nature of the network among all flows [15, 16, 17]. For instance, Hedera [15] detects flows with sizes more than 100MB (10% of the link's capacity) and estimates their demands based on max-min fairness criterion to reroute them. However, as recent transport designs show, minimizing FCT in DCNs should be done through considering the prioritization in the network. Therefore, following the fairness criterion for designing the load balancing block will cause suboptimal FCT, though a better load balancing design, such as [16, 17], could reduce the overall FCT.

## III. MOTIVATIONS & DESIGN DECISIONS

**Scheduling Over Single-Link vs. Multiple-Link Paths:** It is usually mentioned in the literature that preempting lower priority flows to serve higher priority ones minimizes the AFCT. This statement is a direct result of considering SRPT–the optimum solution when scheduling over a single link–as main algorithm to schedule flows (e.g., [1, 2, 3]). However, we show that this statement is wrong in a network where paths contain multiple links. For that purpose, we use a simple example shown in Fig. 1 where flows #1 and #2 have 5 and 4 remaining units respectively. Now a new flow (Flow #3) with 3 units comes to the network (consider remaining size of each flow as its priority i.e., smaller size has higher priority). So clearly, in contrast with SRPT, using no preemption (Fig. 1.a) leads to smaller AFCT. This is important to mention that using either local-aware SRPT (in S1 and S2 switches) (as in [1]) or global-aware SRPT (as in [3]) will lead to the suboptimal result (Fig. 1.b). Therefore, the incorrectness of the mentioned statement illustrates the need for designing better scheduling algorithms by considering the multiple-link nature of paths in DCNs.

**Simple, deployable, and end-to-end:** Datacenter owners usually prefer using scale out (using commodity switches) to scale up (using high-end switches with high-end new features) to build their networks [18, 7]. This motivates us to not modify any switches in the network, though modifying switches might give good performance [1, 16, 6, 24] and look for a simple end-to-end solution which is deployment friendly.

**Why Hybrid?** Centralized approaches are attractive because they could use global knowledge of the network to make better decisions [15, 8]. However, they suffer from some issues. Due to the communication delay with the controller, scheduling small flows (most flows in DCNs [9]) through centralized approaches is not desired. Another issue is their response times. For instance, the scheduler in [15] runs every 5 seconds, which leads to its bad performance compared to distributed solutions such as [16]. For centralized schemes such as [8] that require highly synchronized nodes, synchronization is another issue. Keeping nodes synchronized at the order of one microsecond as [8] requires, is challenging in a real DC environment [5]. On the other hand, although responsiveness of distributed approaches [16, 6, 2] is good, they require adding new functionalities to the switches. Therefore, instead of using a fully distributed or a fully centralized technique, it is beneficial to come up with a hybrid approach combining the global awareness of centralized techniques and the agility of distributed ones.

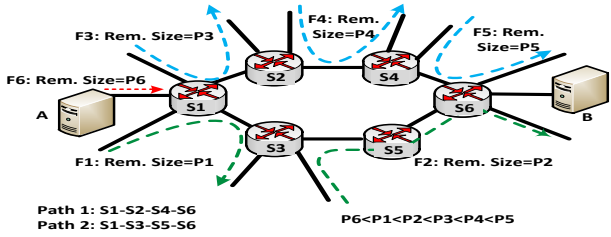

Fig. 2. A simple multipath network.

**Why Path-Aware Scheduling?** One of our main ideas is that load balancing and flow scheduling are dependent design blocks in DCNs and should be designed together to achieve a global objective such as minimizing AFCT in a multipath DCN. So, instead of using single-path scheduling policies (e.g., [1, 4, 6, 5, 2, 3, 24]), we consider a path-aware scheduling logic.

## IV. DESIGN

Scheduling flows to minimize AFCT in single path scenario is an NP-hard problem [1]. This problem in multipath scenarios will remain NP-hard. In this section, we introduce the key design principles of HyLine, which uses heuristic approach to minimize AFCT using path-aware scheduling in multipath commodity DCNs.

### A. HyLine's Big Picture

**End-Hosts:** In HyLine, end-hosts are responsible for classifying all flows into two classes: 1) Latency-sensitive flows, i.e., the small flows, that require less queuing and transmission delays. 2) Bandwidth hungry flows that could tolerate some delays during their transmission. This classification will be done using a threshold provided by MAN, a logically centralized network manager. All of the flows in the latency-sensitive class (1st class i.e., flows having sizes smaller than threshold) are assigned to the higher priority queue in switches ($Q_1$) and all of the bandwidth hungry flows (2nd class i.e., flows having sizes bigger than threshold) are assigned to the lower priority queue in switches ($Q_2$). Next, all 1st class flows are sent to the network at line rate, and flow-based ECMP is used for balancing their loads among available paths. However, end-hosts should first send a Request to Send (RTS) message to MAN asking permission before sending any of their 2nd class flows to the network. This RTS carries source, destination, and size of the flow.

**MAN:** MAN is the logically centralized entity in HyLine that is responsible for scheduling 2nd class flows. To this end, it guides transmission of all of the 2nd class flows in a very simple Stop-and-Go fashion. If MAN decides that a flow could come to the network, it sends back a Clear to Send (CTS) message (i.e., Go) carrying the path that should be used for transmission of this flow. If not, it sends back a Stop to Send (STS) message forcing the flow to be kept at the edge of network. Flows that get CTS messages are sent to the network at line rate. These permitted flows only would be stopped momentarily in two conditions by two different mechanisms:

*First:* When there is no more bandwidth available to serve a new incoming 2nd class flow with higher priority than a few of the permitted ones. In this case, MAN uses a path-aware preemption mechanism (§4.3) to select the best set of flows to preempt and sends the STS messages to the preempted ones and stops them.

*Second:* When permitted 2nd class flows are going to be dropped at switches due to a high load in higher priority queue caused by 1st class flows. In this case, to keep the design simple and practical, instead of using fine-grained monitoring of the queue occupancies for each switch, PFC– defined as part of IEEE 802.1Qbb standard [19] and an available feature in today's commodity switches [9, 20]–is used to pause permitted 2nd class flows without any need for coordination with MAN.

When a 2nd class flow is finished (or close to being finished), its corresponding end-host sends a FIN message to MAN indicating that the path (and bandwidth) allocated for this flow is now free. Then, MAN assigns the available resources to other flows which are stopped (by MAN).

### B. Why it works?

There are three main reasons why HyLine boosts performance of latency sensitive flows in DCNs:

1) Queue length builds up in a DCN mainly as a result of having bandwidth hungry flows. This class of flows occupies queues and causes dramatic increase in completion times of small flows due to increasing buffer delay and increasing drop rate of small flows' packets and the consequent retransmission of them. Therefore, giving credit to small flows and allowing them to be served first in the switches significantly reduces their completion times.

2) Due to the hash-based nature of flow-based ECMP, this load balancing scheme performs very well when it is used for a network that consists of only small flows [16].

3) Making the bandwidth hungry flows (large portion of all bytes transferred in DCN [9, 7]) to be served after serving the 1st class flows opens room for the 1st class flows to bypass the slow start phase of TCP and finish as soon as possible.

In addition, HyLine boosts performance of bandwidth hungry flows, i.e., the 2nd class compared to single-path based flow scheduler proposals [9, 1, 4, 5], because:

1) Using the MAN, a logically centralized network manager, enables HyLine to have global knowledge of the network for scheduling the 2nd class flows.

2) HyLine benefits from the pre-planned nature of DCN topologies and uses a preemption policy that not only considers flows' information but also network's topology information at the time of scheduling.

3) Since HyLine pushes back and stops the 2nd class flows at the edge of the network when network could not serve them at the current time, packet drops, retransmissions, queue occupancy, and congestion for the 2nd class flows are reduced dramatically.

### C. Path-Aware Flow Scheduling Heuristic

In this section, we introduce a new path-aware scheduling policy used in the core of HyLine by considering multiple-path DCNs where each path has multiple links.

**To Preempt or Not to Preempt:** To explain the HyLine's path-aware scheduling policy, we use the example shown in Fig. 2. Flow #6 (F6) with size $p_6$ is generated at A and destined to B, while there is no enough bandwidth to serve this flow without preempting others (different links might contain different flows, but Fig. 2 only shows the ones that have lower priorities (higher remaining sizes) than F6). Similar to the example in Fig. 1, total flow completion time (TF) when using each path can be calculated as follow:

Without Preemption:

$$\begin{cases} TF_{Path1} = [p_1 + p_2] + [p_3 + p_4 + p_5 + (p_6 + p_5)] = \sum_1^6 p_i + p_5 \\ TF_{Path2} = [p_1 + p_2 + (p_6 + p_2)] + [p_3 + p_4 + p_5] = \sum_1^6 p_i + p_2 \end{cases} \quad (1)$$

With Preemption:

$$\begin{cases} TF_{path1} = [p_1 + p_2] + [(p_6 + p_3) + (p_6 + p_4) + (p_6 + p_5) + p_6] \\ \qquad\qquad = \sum_1^6 p_i + 3p_6 \\ TF_{Path2} = [(p_1 + p_6) + (p_2 + p_6) + p_6] + [p_3 + p_4 + p_5] \\ \qquad\qquad = \sum_1^6 p_i + 2p_6 \end{cases} \quad (2)$$

As these equations illustrate, path 2 is the best choice, and if $2p_6 < p_2$, preemption should be used.

In general, when N, $P_{max}$, and $P_{new}$ represent number of required preemption on a path, maximum priority on a path, and priority of the new flow, if $N \times P_{new} < P_{max}$, preemption is preferred, while in other cases, using no-preemption leads to smaller AFCT. Therefore, totally, the path that has the Minimum of either $N \times P_{new}$ (in short, MNP) or $P_{max}$ is the best path.

### D. Scheduling Logic's Details

HyLine's main path-aware scheduling policy is based on the fact that permitted flows are sent at edge link's line rate. This makes the overall design very simple and omits the need for any precise rate calculation and sophisticated scheduling policies. Another key rule to simplify the logic and reduce the time complexity is out-of-order delivery avoidance. To avoid out-of-order delivery, paths allocated for permitted flows could not be changed. In other words, only new flows and already stopped ones (by MAN) could be assigned to other paths.

Algorithm 1 shows MAN's main logic. With new incoming (RTS) request for a flow, MAN looks for the best path for the new flow. For this purpose, MAN finds the number of required preemptions and lowest priority on each path.

**Balanced Load:** When new flow is permitted to come to the network, and there are multiple choices for the final path, the remaining BW of these paths is considered and the path with the maximum remaining BW is selected for the new incoming flow (Remaining BW of a path is defined as the minimum remaining BW of the links in that path). If the remaining BW is also equal for those paths, random selection will be used to break the tie. MAN will only consider 2nd class flows to calculate remaining BW, because it does not have any information about 1st class flows. HyLine manages the impact of 1st class flows by using PFC in the network.

---

**Algorithm 1:** HyLine's Main Algorithm

1  **Function** *NewRequest(f)*
2     $FlowList \longleftarrow f$ /* insert to the sorted list   */
3     $Schedule(f)$;
4  **Function** *RemoveRequest(f)*
5     $RemoveFlowfromPath(f.path, f)$
6     $FlowList \longrightarrow f$ /* remove $f$ from list          */
7     $ReSchedule()$;
8  **Function** *Schedule(f)*
9     $[found, premptList, path] = FindPath(f)$
10    **if** $found$ **then**
11       **for** $Flow \in premptList$ **do**
12          $RemoveFlowfromPath(path, Flow)$
13          $SendSTS(Flow.src)$
14       $AddFlowtoPath(path, f)$
15       $SendCTS(f.src, path)$
16       $f.IsStopped = false; ReSchedule()$
17       **for** $Flow \in premptList$ **do**
18          $Flow.IsStopped = true$
19    **else**
20       $SendSTS(f.src)$
21 **Function** *ReSchedule*
22    **for** $f \in Flowlist$ && **if**$(f.IsStopped)$ **do**
23       $(found, premptList, path) = FindPath(f)$
24       **if** $found$ **then**
25          **for** $Flow \in preemptList$ **do**
26             $RemoveFlowfromPath(path, Flow)$
27             $SendSTS(Flow.src)$
28             $Flow.IsStopped = true$
29          $AddFlowtoPath(path, f)$
30          $SendCTS(f.src, path)$
31          $Flow.IsStopped = false$
32 **Function** *FindPath(f)*
33    $mnp = \infty; bw = 0; pathList = void;$
34    $MinMaxPrio = \infty; preemptionList = void$
35    **for** $p \in availablepaths(f.src, f.dst)$ **do**
36       $MinMaxPrio = UpdateMinMaxPrio(p, f, MinMaxPrio)$
37       $found = findMNP(p, f)$
38       **if** $found$ **then**
39          **if** $mnp > p.mnp$ & $p.remBw > bw$ **then**
40             $pathList.clear(); pathList.insert(p)$
41             $mnp = p.mnp; bw = p.bw$
42          **else if** $mnp == p.mnp$ & $p.remBw == bw$ **then**
43             $pathList.insert(p)$
44    **if** $pathList.size \geq 1$ **then**
45       $found = true; path = pathList.Random()$
46       $preemptList = p.preemptList$
47    **if** $found$ & $mnp \times f.priority \geq MinMaxPrio$ **then**
48       $found = false$
49    **return** $found, premptList, path$

---

**Reschedule:** After selecting a path for a new incoming flow and likely stopping/preempting some other flows on this path, there might be available room for flows that have been stopped before. Therefore, in case of preemption, MAN checks the possibility of admitting more flows into the network (considering out-of-order delivery avoidance rule). Clearly, there is a trade-off between adding more rounds of rescheduling to admit more probable flows and the overall time complexity of the algorithm. To reduce the time complexity of the main logic, we decided to do only one round of rescheduling. The results in §5 show that this decision still leads to very good overall performance.

**HyLine's Time Complexity:** Here, we show that the time complexity of HyLine is O(|F|) where |F| is the total number of active flows in the 2nd class. To show this result, we first should notice that the maximum number of permitted flows on a link has an upper bound that is independent of the number of flows

considering the assumption that all flows are sent at line rate. Assuming that the lowest and highest link rates on a path are S bps and M bps, respectively, the maximum number of 2nd class flows in a link of that path is M/S. When findMNP procedure (line 37 in Algorithm 1) is implemented simply by exploring the entire valid preemption list of flows in a path, for each path, at most, it looks at $(M/S)^l$ combinations in which, $l$ is number of links in a path. For instance, in a 3-tier datacenter, $l$ is equal to 6. Therefore, FindPath takes constant time. In addition, the out-of-order delivery avoidance rule causes a flow to be considered during the ReSchedule procedure at most once. This illustrates that the Complexity of Schedule function, which is equal to the total complexity of the algorithm, is O(|F|).

**PFC and Head-of-Line Blocking Issue:** PFC if used in a normal network will cause head-of-line blocking issue for flows using the same priority queue. However, in HyLine, MAN pushes most of the 2nd class flows back and stops them from coming into the network. This strategy significantly reduces the head-of-line blocking issue and as the results in §5.5 show, using PFC boosts the overall performance.

**Rate Control:** HyLine has no complicated rate control mechanism. It uses TCP, and to send flows at line rate, changes initial congestion window size. However, 2nd class flows being stopped by MAN should not cause TCP time-outs. So we modify TCP to avoid such time-outs for the 2nd class flows (when they receive STS signal from MAN) without affecting TCP time-out mechanism for 1st class flows. This modification only requires adding a few lines of code to the original TCP implementation.

*E. The Threshold to Distinguish Classes of Traffic*

Some recent studies [1, 4] tried to formalize the problem of finding optimum thresholds to distinguish different flows based on their sizes and use available priority queues in today's commodity switches to separate packets of different types of flows. PIAS and pFabric use simple M/M/1 and M/G/1 queue models, respectively, to find the best threshold values. Even with these simplifications, the problem of finding optimum thresholds is complicated [1] and NP-hard [4]. Moreover, these simplifications do not work in our case. In fact, none of the M/M/1 or M/G/1 FIFO queue models are valid approximations for our 2nd queue. Even in simple single queue scenario where our 2nd queue scheduling mechanism is equal to SRPT, FIFO queue models should be replaced by complex SRPT queue models [21]. From this point of view, the problem of finding the best thresholds becomes even more complicated than before. Therefore, in this paper, we choose another direction and instead of finding the optimum threshold, we determine a band for practical threshold values.

**Lower Bound:** In theory, forcing more flows to be controlled by MAN (i.e., decreasing the threshold) increases the performance because of having a global view of the network during the scheduling; however, in practice, reducing threshold (H) causes additional delays for the small flows due to the controller's delay (both network delay for reaching the MAN and computation delay of MAN). To simplify the analysis and find a lower bound for H, we consider single queue model and use mean queue analysis. We define a delay cost, $T_{cost}$, for any flow which is controlled by MAN, $f_s$ as the smallest flow in the 2nd queue ($f_s$'s size = H), and $W_{f_s}$ as expected waiting time of $f_s$ (time from when it first arrives to when it receives service for the first time). We argue that $W_{f_s}$ should not be smaller than $T_{cost}$ (if $T_{cost} > W_{f_s}$, putting $f_s$ in the 1st queue (increasing the threshold to $H + \varepsilon$) will cause lower FCT for $f_s$).

The $f_s$ will be served only after serving all flows in the 1st queue and after serving all flows having smaller remaining sizes (but originally bigger) than it in the 2nd queue. In other words, any flow with smaller size than $f_s$ in 1st queue or any flow with smaller remaining size than $f_s$ in 2nd queue will preempt $f_s$ before it receives service for the first time. So, from $f_s$'s point of view, $W_{f_s}$ in this network is equal to $W_{f_s}$ in a network where there is only a single SRPT queue. Therefore, we can use [21]'s analysis for an M/G/1/SRPT queue to find average value of $W_{f_s}$:

$$E[W(x)] = \frac{\lambda\left(m_2(x) + x^2(1-F(x))\right)}{2(1-\rho(x))^2} \quad (3)$$

Here, we denote average arrival rate by $\lambda$, service time (service time=size/service rate, service rate=link speed) of a flow by X, CDF of service time distribution by $F(x)$, $m_2(x) = \int_0^x t^2 f(t)dt$, and the load made up by the flows of service time less than or equal $x$ by $\rho(x) = \lambda \overline{X_x}$ in which $\overline{X_x} = \int_0^x tf(t)dt$. Substituting $x$ in this formula with $h = H/C$ in which C represents bandwidth of the link, will lead to calculation of $E[W(h)] = E[W_{f_s}]$. So, the following inequality represents the lower bound:

$$T_{cost} \leq \frac{\lambda\left(m_2(h) + h^2(1-F(h))\right)}{2(1-\rho(h))^2} \quad (4)$$

**Upper Bound:** Increasing H puts more flows into the 1st class, causes congestion in the 1st queue and consequently decreases the performance. Therefore, to address this issue, we require an upper bound on H. Here, the important observation is that almost all of the schemes including normal TCP perform very well when load is very low (less than 10%) [1, 4, 3, 5]. The reason is that at low load, inter-arrival of the flows is large enough to serve flows without having congestion issue. Based on this important observation, we cap the overall load of the 1st queue. In more detail, we choose $\rho_1 = \rho(h) \leq \frac{\rho_{total}}{10}$. Since $\rho_{total} < 1$, this choice guarantees that the total load in the 1st queue ($\rho_1$) is always smaller than 10%; therefore, congestion in the 1st queue will not be an issue. So, following equation represents the upper bound:

$$\frac{\rho_1}{\rho_{total}} = \frac{\overline{X_h}}{\overline{X_{total}}} \leq 0.1 \quad (5)$$

$E[W(H)]$ and $\frac{\rho_1}{\rho_{total}}$ for web search workload and different loads (up to 90%) are shown in Fig. 3 (C=1Gbps). The band for choosing H in a moderate load of 60% is depicted in this figure too (through this paper we assume $T_{cost} = 100\mu s$).

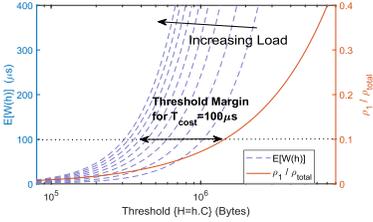
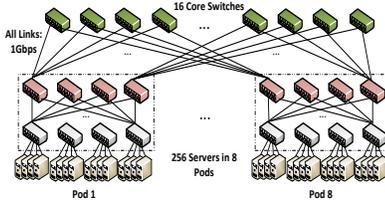
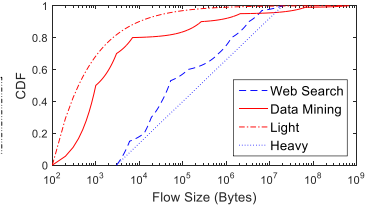

Fig. 3. $E[W(H)]$ and $\rho_1/\rho_{total}$

Fig. 4. Simulation setup (a) The 3-tier topology (b) Flow size distributions of workloads used

**Static vs. Dynamic Threshold Assignment:** Clearly, assigning thresholds dynamically based on the load of the flows (as lower bound criterion suggests) is beneficial. For that purpose, different agents at end-hosts could periodically report summary of all their flows to the MAN. Later, MAN will use these reports to choose the threshold and report it back to the end-hosts. Although HyLine's structure enables us to use this approach, to keep the design simple and practical we use a static threshold assignment, and in §5, we will show that this approach works very well for different loads and even for different types of workloads. So, through the rest of this paper, based on Fig. 3, we choose H=1MB.

## V. EVALUATION

In this section, we evaluate the performance of HyLine using extensive packet-level simulations in ns2 (available at: https://github.com/soheil-ab/hyline). First, we compare the performance of HyLine with existing proposals including Qjump [5], pFabric [1], DCTCP [9], and TCP-New Reno. Then, through micro-benchmarks, we evaluate HyLine's performance such as its sensitivity to the threshold value, improvements caused by PFC.

### A. Simulation Settings

**Datacenter Topology:** We use a 3-tier fat-tree topology [18] which is the base topology for today's DCNs [22,13] for our evaluation (Fig. 4.a). The topology includes 8 pods interconnecting 256 end-hosts using 80 8-port switches with a 300μs overall end-to-end RTT delay between end-hosts located in different pods.

**Load-balancing Mechanism:** To have a fair comparison of HyLine's performance with other single-path based flow-scheduling schemes, we use flow-based ECMP used in commodity DCNs [18, 7] as the load balancing scheme.

**Traffic Workloads:** We use two realistic workloads from production datacenters: web search workload [9] and data mining workload [7]. In addition, we use 2 other synthetic workloads named Heavy and Light to change the heavy-tailedness of the traffic and do stress tests. The flow size distributions of all workloads are shown in Fig. 4.b.

**Performance Metrics:** We consider AFCT and 99th percentile FCT of flows as the performance objectives like prior work [1, 4, 2, 3]. We normalize all FCTs to the flows' ideal values achieved if each flow is transmitted over the fabric without any interference from competing traffic. In addition, since most of the datacenter applications (from search and social networks to MapReduce) use partition-aggregate structure equipped with different deadlines for flows in different layers of its hierarchy [23], similar to prior work [1, 3, 23], we use the application throughput defined as the fraction of flows that meet their deadline as another performance metric to investigate the impact of HyLine on real applications.

**Schemes Compared:** We compare HyLine with Qjump [5], pFabric [1], DCTCP [9], and TCP-New Reno with Sack. The parameters used for the evaluation of these schemes are selected based on their authors' recommendations or reflect the best settings that we have experimentally determined (Table 1). We use these parameters for evaluations in this section unless otherwise specified.

**PFC Implementation in ns2:** We use a simplified version of PFC (on/off style) that we have added to ns2 simulator. For that purpose, when the queue size hits a threshold (pause threshold), the switch sends pause signal to upstream switch. When the queue size becomes less than another threshold (resume threshold) the switch sends resume message.

### B. Overall Performance

In this section, we present the overall performance of HyLine under the aforementioned workload and DCN topology. [9, 7]. We show that despite HyLine's simplicity, it outperforms all compared schemes.

*Overall AFCT:* The overall normalized FCT of flows with different schemes for search and data mining workloads are shown in Fig. 5.e and Fig. 6.e, respectively. As these results illustrate, HyLine achieves the best performance among all compared schemes. For instance, AFCT using HyLine is ~3-31% and ~52-66% lower than pFabric and Qjump respectively. All schemes generally perform better in data mining workload. The reason is that in this workload probability of having two large flows competing for the same link is less than search workload (Fig. 4.b). For this workload, HyLine achieves ~18-30% lower AFCT than Qjump and compared to pFabric performs roughly the same.

*AFCT in More Detail:* As expected, pFabric performs well for the very small flows in (0, 100kB] range (Fig. 5.a and Fig.

TABLE I. DEFAULT SIMULATION SETTINGS

| Scheme | Parameters |
|---|---|
| pFabric | qsize = 50pkts (=2×BDP), initCwnd = 25pkts (=BDP), minRTO =1ms (≈3× RTT) |
| Qjump | qsize = 225pkts, initCwnd = 25pkts, minRTO = 4ms |
| dctcp & tcp | qsize = 225pkts, initCwnd = 10pkts, minRTO = 4ms |
| HyLine | qsize = 225pkts, initCwnd = 25pkts, H=1MB, Tcost=100us, minRTO = 4ms, initCwnd (2nd class) = 25pkts, minRTO (2nd class) = 1s, pause threshold=215pkts, resume threshold=205pkts |

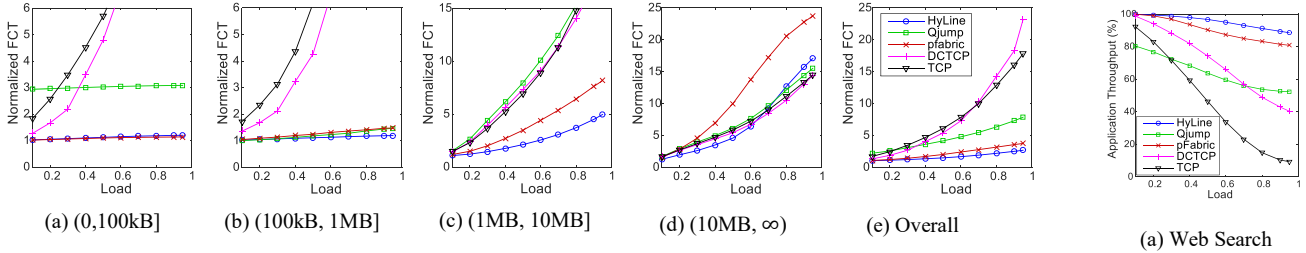

Fig. 5. Normalized FCT statistics across different flow sizes for web search workload.

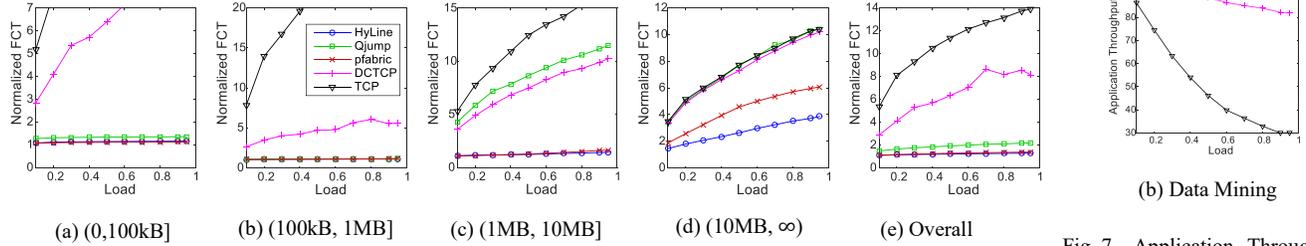

Fig. 6. Normalized FCT statistics across different flow sizes for data mining workload.

Fig. 7. Application Throughput across different loads

6.a). However, it comes at the expense of performance reduction for other ranges of flows, due to its local strategy of dropping packets at earliest stages of the network and reacting to this sooner by using small priority queues in switches and small timeouts at end-hosts. In contrast, HyLine allows the other 1st class packets (i.e. flows in (100kB, 1MB] range) to be queued in switches too. Considering multipath nature of network and the fact that all of these 1st class flows will not compete for the same output links in the next stages of the network, this increases the chance of serving flows in (100kB, 1MB] range later in the network (Fig. 5.b and Fig. 6.b). Moreover, Qjump cannot achieve very good performance for the small flows (specially for the search workload (Fig. 5.a)) because it reduces the throughput of these flows to give more bandwidth to the bigger ones.

For the 2nd class flows ((1MB, 10MB] and (10MB, ∞) ranges), HyLine benefits from having a global view and path-aware nature in its scheduler compared to other schemes . So, as Fig. 5.c-d and Fig. 6.c-d illustrate, it performs better than all other schemes for almost all loads and workloads except very high loads in search workload for flows in (10MB, ∞) range. For high loads in this range (Fig. 5.d), since total number of flows including big flows increases, the total number of preemptions for this range of flows increases too. Therefore, largest flows in the network face more preemption delay. In contrast, TCP achieves best performance at high loads (Fig. 5.d), because it loses less bandwidth due to the fairness nature of its design.

### C. Varying Performance Metrics

**Application Throughput:** Most of the today's datacenter applications use partition-aggregate structure in which flows in each level of the hierarchy have deadlines [23]. For instance, in a search application, if responses (flows) from workers miss their deadlines, they are not included in the total response, typically hurt the response quality, and waste network bandwidth. Therefore, to investigate impact of HyLine for such applications, we assign different deadlines to different flows and similar to prior work [1, 2, 23] consider application throughput as the performance metric. Here, deadline of each flow is considered 4x of its ideal completion time achieved when there are no other competing flows in the network. We used tighter and looser deadlines for flows too, but since the overall results are similar to the presented results, for brevity, we only report the results for the mentioned deadline. Fig. 7 depicts the overall results for two realistic workloads across different loads. HyLine outperforms other schemes for both workloads.

Since in both workloads, most traffic are small flows, finishing these small flows faster increases the probability of meeting their deadlines. Therefore, schemes which achieve better results for small flows potentially perform better for deadline-aware traffic too. That's why HyLine and pFabric perform very well compared to other schemes. It is important to notice that HyLine achieves this performance without any changes in the network, while pFabric requires changes in switches.

**99th Percentile:** In addition to previous metrics, we also consider the 99th percentile FCT as a performance metric to have a better comparison of HyLine with other schemes. Fig. 8 and Fig. 9 show the results of 99th percentile FCT for data mining and search workloads respectively for different flows' size ranges. $99^{th}$ percentile result's pattern is similar to the AFCT result's pattern discussed earlier.

### D. Impact of Workload

So far we evaluated HyLine under realistic DCN workloads. However, there might be still two concerns about the HyLine's performance: 1-What if traffic consists of more 1st class flows? 2-What if a workload consists of more 2nd class flows? To evaluate the performance under these two corner cases, we used Bounded-Pareto distribution to generate 2 synthetic workloads named Light and Heavy (Fig. 4.b). In Light workload, 97% of the flows are smaller than 100KB, while this number is only 40% for Heavy workload. This will provide us with workloads to check the two mentioned concerns. Fig. 10 and Fig. 11 show the AFCT, and application throughput results using Light and Heavy workloads. Under Light workload, all schemes generally

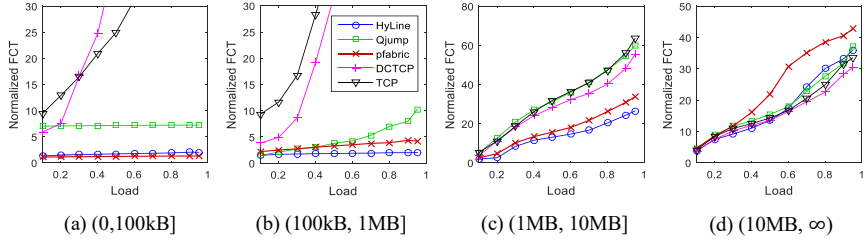
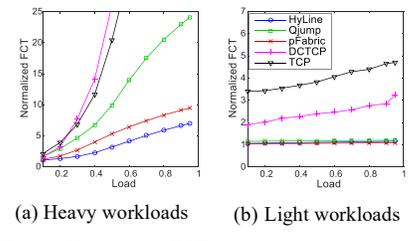

(a) (0,100kB)   (b) (100kB, 1MB]   (c) (1MB, 10MB]   (d) (10MB, ∞)

Fig. 9. Normalized 99th percentile FCT statistics for web search workload across different flow sizes.

(a) Heavy workloads   (b) Light workloads

Fig. 10. Normalized FCT statistics

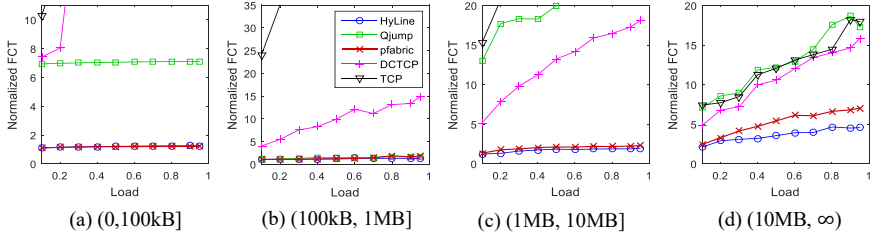
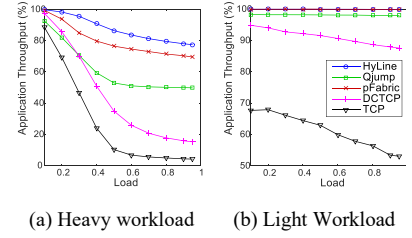

(a) (0,100kB)   (b) (100kB, 1MB]   (c) (1MB, 10MB]   (d) (10MB, ∞)

Fig. 8. Normalized 99th percentile FCT statistics for data mining workload across different flow sizes.

(a) Heavy workload   (b) Light Workload

Fig. 11. Application Throughput

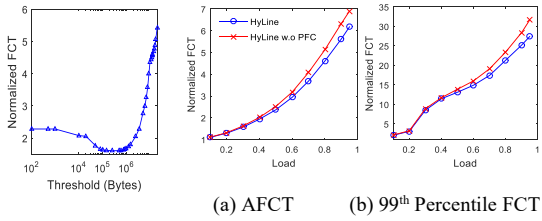
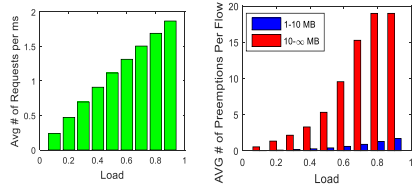
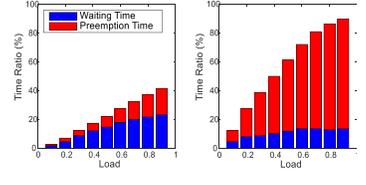

(a) AFCT   (b) 99th Percentile FCT

Fig. 12. AFCT across different thresholds.

Fig. 13. Performance with & without PFC

Fig. 14. Average # of req. received by MAN

Fig. 15. Avg. # of preemptions per flow for 2 ranges of flow sizes

(a) (1-10MB) range   (b) (10MB-∞) range

Fig. 16. Average portion of waiting and preemption times that a flow experiences to the total flow's completion time.

perform well. However, for Heavy workload including more big flows, the performance of schemes drops dramatically. Here, scheduling issue and handling big flows dominate, and the scheme which manages these issues better than others will achieve higher performance. That's why compared to other schemes, HyLine works very well under Heavy workload.

*E. HyLine Deep Dive*

In this section, a series of targeted simulations are conducted to dig deeper into HyLine's design.

**Sensitivity to Threshold:** To check our analysis in the §4.5, we use search workload and change the threshold identifying the two classes of traffic, and check the AFCT as the performance metric. Fig. 12 presents overall results in 60% load. Here, the results fit very well with our lower bound and upper bound analysis (Fig. 3). As we expected, for the thresholds below the lower bound, cost of doing centralized scheduling dominates, and for the ones above the upper bound, benefits of using centralized scheduler is not so much. So, in both cases, overall performance drops.

**PFC:** PFC, if used in a normal network, could cause the head-of-line blocking issue. However, since HyLine controls all of the 2nd class flows in the network, it prevents the head-of-line blocking issue for this class of flows. Moreover, PFC is used to prevent any drop of the 2nd class packets due to the increase in the number of 1st class flows at high load situations. To show the impact of using PFC at high loads, we use web search workload and do simulations with and without PFC feature in switches. Fig. 13 illustrates the improvement of the overall performance for the 2nd class of flows when PFC is turned on. In fact, PFC improves AFCT and 99th FCT by up to 13% and 15% respectively at high loads.

**MAN:** Here, we report MAN's performance measurements including average number of requests that MAN receives (Fig. 14), average waiting time (the time from when a flow first arrives at the end-host to when it receives first CTS (GO signal) from MAN), average preemption time (the total time that a flow is in STOP state (i.e., preempted by MAN)), and average number of preemptions that a flow experiences under web search workload. When load increases, as expected, number of preemptions per flow for the biggest flows (in (10MB-∞) range) increases (Fig. 15). However, since smaller flows (in (1-10MB] range) could be finished faster due to no competing bigger flows which are already stopped by MAN at the edge of network, the probability of being preempted during their transmission will be small. This is shown in Fig. 15.

Fig. 16 shows ratio of waiting and preemption times that on average a flow experiences to its total completion time across different loads for 2 different ranges of flow sizes. As mentioned earlier preemption time of flows in (1-10MB] range is small. Also, as loads increases waiting time of flows in this range slightly increases. The reason is that the flows which already have got permission from MAN most likely have smaller remaining sizes compared to the new incoming flows, so new incoming flows will wait for the completion of these flows.

## VI. DISCUSSION

**Flow Information:** Previous studies show that for many DCN applications (e.g. web search, Hadoop [25], data processing), size of the flows are known at initiation time (For example, see §2.1 of [23]), and can be conveyed to lower layer (e.g., through a socket option). In other cases, when sizes of flows are not known precisely in advance, offline measurements enable applications to have an approximation of the flow sizes and use them later at run time. However, it is important to mention that based on DCN's traffic characteristics, HyLine does not require exact size information for 1st class flows (most of the DCN's flows), because it let them come to the network without scheduling them one by one, while all other size-aware schemes (e.g., [1], [2], [3]) need to know the exact size of all of the flows. Therefore, using HyLine, for most of the DCN's flows, these offline measurements will be just to check whether a flow is less than a threshold (e.g., 1MB).

**Stopping vs. Terminating an Application:** Although results shown in Fig. 16 indicate that most of the 2nd class flows have a very small preemption time, it is worth mentioning that stopping a 2nd class flow momentarily is not equal to terminating it. From the applications' point of view, it is more like TCP being in slow phase, so when a flow is stopped momentarily by MAN, connections are still there and applications are not terminated. Also, as mentioned before, HyLine modifies TCP to avoid having time-outs in STOP state and reacting to them as indication of packet loss.

**Line Rate Transmission:** With today's advances in both software (e.g., Intel DPDK [26], SR-IOV [27]) and hardware (e.g., [29], [30]), end-hosts can achieve line rate transmissions. However, if applications become the bottleneck of sending at line rate, for the 2nd class flows, they could simply add their maximum capable sending rate (maxRate) as part of their request message to MAN. MAN could consider those flows as flows generated by end-hosts having virtual maxRate-links (instead of their physical speed links). Therefore, without changing the logic, it could allow more flows to come and use same links.

## VII. CONCLUSION

We presented HyLine a simple and practical flow scheduling design for DCNs. HyLine's path-aware scheduling policy exploiting the multipath nature of today's DCNs shows that load-balancing and flow scheduling design blocks are dependent blocks, and they should be designed together to minimize AFCT in DCNs. Moreover, HyLine's hybrid approach indicates that to reach high performance and minimize AFCT, it is unnecessary to use fine-grained scheduling structures trying to schedule every flow in DCNs by calculating and assigning either precise rates or priorities to them. In sum, Despite the simple nature of HyLine's design and the fact that it does not require any changes in the fabric, our evaluation results show that it outperforms recent flow scheduling solutions. That's why HyLine is a good candidate to be used in today's commodity DCNs, and why we believe that *performance meets simplicity at HyLine*.